\documentclass[12pt]{revtex4}
\usepackage{amsmath}
\usepackage{natbib}
\usepackage{graphicx}
\usepackage[T1]{fontenc}
\usepackage{times}
\newcommand{\opA}{\hat{\mathrm{A}}}
\newcommand{\opB}{\hat{\mathrm{B}}}
\newcommand{\vB}{\mathbf{B}}
\newcommand{\opC}{\hat{\mathrm{C}}}
\newcommand{\ud}{\mathrm{d}}
\newcommand{\mD}{\mathrm{D}}

\newcommand{\e}{\mathrm{e}}
\newcommand{\vE}{\mathbf{E}}

\newcommand{\mG}{\mathrm{G}}
\newcommand{\opG}{\hat{\mG}}
\newcommand{\opg}{\hat{\mathrm{g}}}
\newcommand{\opGG}{{\overline{\mG}}}
\newcommand{\opgg}{{\overline{\mathrm{g}}}}
\newcommand{\ii}{\mathrm{i}}
\renewcommand{\Im}{\mathrm{Im}\,}

\newcommand{\opI}{\hat{\mathsf{1}}}
\newcommand{\mI}{\mathrm{I}}

\newcommand{\mO}{\mathrm{O}}
\newcommand{\oprho}{\hat{\varrho}}
\newcommand{\opSigma}{\hat{\Sigma}}
\newcommand{\opO}{\hat{\mO}}
\newcommand{\mP}{\mathrm{P}}
\newcommand{\vp}{\mathbf{p}}
\newcommand{\vq}{\mathbf{q}}
\newcommand{\up}{\hat{\vp}}
\newcommand{\mQ}{\mathrm{Q}}
\renewcommand{\vr}{\mathbf{r}}
\newcommand{\mR}{\mathrm{R}}

\newcommand{\vPsi}{\mathsf{\Psi}}
\newcommand{\SO}[1]{\mathit{SO}(#1)}

\newcommand{\mT}{\mathrm{T}}
\newcommand{\opT}{\hat{\mT}}
\newcommand{\opTT}{\overline{\mT}}

\newcommand{\mX}{\mathrm{X}}
\newcommand{\mY}{\mathrm{Y}}
\newcommand{\mZ}{\mathrm{Z}}

\newcommand{\uz}{\hat{\mathbf{z}}}
\DeclareMathOperator\thop{\text{\dh}}

\newcommand{\Ket}[1]{\left|{#1}\right\rangle}
\newcommand{\lmn}[3]{\begin{smallmatrix}{#1}\\{#2}{#3}\end{smallmatrix}}
\newcommand{\Bra}[1]{\left\langle{#1}\right|}
\newcommand{\Bracket}[2]{\left\langle{#1}\,\vrule\,{#2}\right\rangle}
\newcommand{\BraOpKet}[3]{\left\langle{#1}\,\vrule\,{#2}\,\vrule\,{#3}\right\rangle}
\newcommand{\deltar}{\delta_\perp}

\begin{document}
\title{A general framework for multiple scattering of polarized waves
including anisotropies and Berry phase}
\author{V. Rossetto}
\email{vincent.rossetto@grenoble.cnrs.fr}
\affiliation{Universit\'e Joseph Fourier, Grenoble - FRANCE\\
             LPMMC - CNRS %
             Maison des Magist\`eres - BP 166 \\%
             25, avenue des Martyrs - 38042 Grenoble CEDEX - FRANCE}

\begin{abstract}
We develop a framework for the multiple scattering of a polarized wave.
We consider particles with spin propagating in a medium
filled with scatterers. We write the amplitudes of each spin
eigenstate in a local, mobile frame. One of the axes is in the direction 
of propagation of the particle. We use this representation to
define a \emph{directional} Green's operator of the homogeneous medium
and also to write the spin-dependent scattering amplitudes.
We show that this representation reveals a Berry phase.
We establish a generalized Green-Dyson equation for the multiple scattering
problem in this framework. 
We show that the generalized Green-Dyson equation can be 
solved by linear algebra
if one uses a representation of the rotations based on Wigner D-matrices.
The properties of light scattering are retrieved if we use spin~1 particles.
Our theory allows to take into
account several kinds of anisotropies like circular or linear dichroism
and birefringence, Faraday effects and Mie scattering within the same 
formalism. Several anisotropies can be present at the same time.
\end{abstract}

\maketitle

\section{Introduction}
In optics, several phenomena involve the polarization of light.
Such phenomena, like birefringence, dichroism and Faraday effect, 
are well understood and described in textbooks~\cite{bornwolf}, 
usually in the case where the medium is homogeneous and linear.
When filled with particles, the medium is no longer homogeneous
and in addition to the transport of light, one has to
study its scattering by the particles. A century ago, Mie
discovered that the scattering of light by small dielectric
spheres is linear and also affects the polarization~\cite{mie1908,jackson}. 
The theory of light scattering has been extended
for several different particles and studied in the context of 
multiple scattering~\cite{vandehulst1957,newton,mischchenko}.
While the polarization is well
studied for a single scattering, the problem becomes more
difficult in multiple scattering and a scalar approximation
is often used. Such an approximation makes it simpler to 
investigate the coherent effects using sophisticated
diagram techniques~\cite{vanrossum1999,akkermans}.
As a consequence, polarization effects are neglected
whereas they may, in certains situations, be of interest.
In magneto-chiral birefringence, for instance, a residual degree of
circular polarization persists is chiral media~\cite{pinheiro2003}.
Recent experiments of coherent backscattering of light by cold atoms 
exhibit polarization-dependent enhancement factors~\cite{labeyrie1999};
this result is interpreted thanks to the quantum, nonlinear, nature 
of the cold atoms~\cite{jonckheere2000}. Under certain conditions,
it was even forecast that an anti-localization effect could be
observed~\cite{kuprianov2006}.
Quantum effects of light were also introduced, in second quantization,
for the study of dense cold atomic gases~\cite{sokolov2009}.

Remarkably, most of the scalar approximation techniques are applicable
not only to light but also to microwaves, ultrasound and non interacting 
electrons, as emphasized in Ref.~\cite{vanrossum1999}. 
The existence of a polarization, alike, 
is not restricted to light; other waves display polarization effects:
The other electromagnetic waves, elastic waves, electrons. More
generally every particle with a spin is, from a quantum point of view,
a polarized wave.
The classical light polarization is a consequence of the vectorial nature 
of the electromagnetic field whereas in the quantum-mechanical picture,
photons have two states called helicities. The origin of the helicity
of a particle follows from the spin. Relativity
imposes that the spin~0 state of massless particles does not exist
in the direction of propagation,
photons therefore posses only two helicities.
The freedom of the relative phase between the two helicity states is 
at the origin of the phenomenon of polarization as 
we experience it in optics.  
All quantum particles with a non-zero spin also 
have helicities and hence may display polarization effects.
Therefore, the problem of polarization transport in multiple
scattering does not only arise in optics: 
Polarization transport in multiple scattering is a more general problem
and optics can be seen as the most investigated particular case.

According to the quantum
duality of particles and waves, we may describe the propagation of 
particles as wave amplitudes from a statistical, or probabilistic,
point of view. 
Reciprocally, the propagation of a wave can be described
by the statistics of particle paths, through techniques like Feynman's
path integration.
Following this picture, many theoretical calculations 
related to the multiple scattering of polarized waves are performed by 
Monte-Carlo techniques. Such techniques have been developped for light,
even in the presence of quantum scatterers~\cite{labeyrie2004}, and
for elastic waves~\cite{margerin2000}. Monte-Carlo techniques are
a pragmatic solution to the problem of 
taking into account each trajectory as a whole. Is it possible
to consider each trajectory entirely and perform statistics in
a more formal way ?
A pioneering work on trajectory statistics was made by Sato
\cite{sato1995} with the introduction of a directional Green's
function for the elastic energy (equivalent to the optical intensity)
in a three dimensional medium with
nonisotropic scatterers. However, as his study concerned
energy, polarization was not included in the model. 

The trajectories of energy, or intensities, have been studied
in a multiple scattering context using 
Stokes parameters. These parameters are preferentially used by 
experimentalists because they are, as intensities, directly measurable 
quantities. They depend on the choice of the 
reference frame used for their observation. 
Their ability to take
coherence effects into account is limited to intensity correlations.
In order to dispose of the reference frame indermination,
Ku\v{s}\v{c}er and Ribari\v{c} introduced harmonical functions
to describe the Stokes parameters \cite{kuscer1959}.
Based on Wigner's work concerning group theory~\cite{wigner1959}, these
functions have also been used under the name ``spin-weighted spherical 
harmonics'', though with a different phase factor~\cite{newman1966}.
They are also implicitely used in light scattering studies under the name
``vector spherical harmonics''~\cite{newton,jackson}.
The seminal work of Ku\v{s}\v{c}er and Ribari\v{c} motivated
a large number of studies on the phase matrix~%
\cite{domke1975,siewert1982,lacoste1998}
and on the radiative transfer equation 
\cite{hovenier1983,garcia1986}. 

The role of trajectories appears in 
a surprising effect involving polarized light: A four-foiled pattern is 
observed in 
backscattering experiments of polarized light from a medium containing
anisotropic scatterers~\cite{pal1985,hielscher1997,cameron1998}.
In these experiments, the source is not a planar, infinite wave, 
but a \emph{localized beam}. The effect does not exist if the source is
extended on a scale of the order of the four-foiled patterns. The same medium
does not exhibit such patterns if the observations are made in transmission
rather than in reflection, so that 
the properties of polarization depend not only on the detailed
trajectory, but also on the direction of observation, or more precisely
on the \emph{frame} of observation. It has been suggested that the
appearance of these patterns is due to the existence of a Berry phase
for the photon~\cite{rossetto2002b}. 
In quantum mechanics, Berry phases have been studied already
in many situations~\cite{berry1984,shaperewilczek}. 
In optics, the literature is abundant in 
experimental~\cite{tomita1986}
and theoretical considerations~\cite{berry1987b,bhandari1997}.
If light is not guided in an optical fiber, but rather multiply
scattered, the Berry phase still exists~\cite{maggs2001d}.
The probability distribution of the phase depends on the statistics
of the paths followed by the photons.

The purpose of the present article is to introduce a general framework 
for the multiple scattering of any polarized wave, as simply
and as pedagogically as possible. We will base our construction
on the concept of path statistics, in the spirit of the Monte-Carlo
calculations, or Feynman's path integral, to account for local 
interactions and path-dependent effects.
We express the formalism for particles
with an arbitrary spin~$S$, although 
we consider only linear media and scatterers
and thus leave the other quantum properties to
further investigations.
Our objective is not to provide a complete theory for multiple
scattering of polarized waves but rather to set up a framework
in which such a theory could be developed. On that account we
assume the simplest possible point of view and make several assumptions,
that are briefly discussed.

We have organized our work as follows: 
The representation of the field and the geometry of rotations are
discussed in Section~\ref{sec:geometry}. A directional Green's
operator is discussed in Section~\ref{sec:transport}, it describes the 
transport in a homogeneous medium.
Using the same formalism, we introduce the scattering operator
between incoming and scattered state in Section~\ref{sec:scattering}.
These first three sections constitute the elements of the theory.
We demonstrate that our theory takes into account the Berry
phase without requiring any extra tool in Section~\ref{sec:berry}.
We derive a generalized Green-Dyson equation for multiple scattering
of polarized waves in Section~\ref{sec:trajstat} and
we show that a solution of this equation
is obtained by means of the rotational harmonics, introduced in 
Section~\ref{sec:rotation}.
We discuss the particular case of rotational invariant scatterers in a
rotational invariant medium in section~\ref{sec:asym} and conclude.
Fundamental formul\ae{} concerning the rotational harmonics are
provided in appendix~\ref{sec:harmonics}, and some of the several 
notations used for them are listed in appendix~\ref{sec:other}.

\section{Geometry and polarization}
\label{sec:geometry}
We begin the presentation of our work with the introduction of the
geometry elements used throughout this article. Mainly, the
novelty of our work relies on the representation of the state of
waves in local frames. Polarization is introduced as a transformation
property of the wave state under local rotations.
  We express all space coordinates in a fixed reference frame $XYZ$.
Any local frame $X'Y'Z'$ is therefore obtained after a rotation of the
reference frame. The coordinates of the local frame unit vectors, 
written in the reference frame, form a $3\times3$ matrix, 
which the unique rotation matrix mapping $XYZ$
into $X'Y'Z'$ and preserving orientation. The group of rotations 
is $\SO3$. Considering the rotation associated to a frame
is mathematically equivalent to considering the frame itself.
We henceforth only use the description in terms of matrices
for mathematical convenience and use exclusively the $ZYZ$ 
Euler representation of rotation matrices, such that any 
rotation $\mR$ of $\SO3$ is decomposed into 
\begin{equation}
\mR=\mZ(\phi)\mY(\theta)\mZ(\psi),
\label{eq:decompR}
\end{equation}
where $\mZ(\phi)$ is the rotation around the $Z$-axis of angle
$\phi$. Similarly $\mY$ is a rotation around the $Y$-axis.
The decomposition \eqref{eq:decompR} is unique if $0<\theta<\pi$
and $0\leq\phi,\psi<2\pi$. 
For $\theta=0$ or $\pi$, the decomposition is not unique. 

In our discussion, we use the concept of trajectory,
or path of a particle. Studying the propagation of a wave using 
the particle picture would not be fully general if we were
concerned by the behavior of single particles. But as our goal is 
to describe all possible trajectories as a whole, 
it is known that 
this way of addressing the problem of multiple scattering is correct.
As a consequence, we will simultaneously use the wave picture, as
equivalent to the statistics of particle trajectories, and the particle
picture of the propagating wave. The
spin of the particle emerges as an essential ingredient of the 
theory, after geometry considerations.

We study a medium filled with $N$ scatterers and write $\vr_i$ 
for the position of the scatterer~$i$. 
The trajectory of the particle is a succession of displacements between
points $\vr_i$ and changes of direction at the points $\vr_i$. We
call such a change of direction a \emph{scattering} event.
We consider the momentum $\vp_n$ after $n$ scattering events
and decompose it into
\begin{equation}
\vp_n=p_n\;\mP_n\,\uz
\label{eq:decompP}
\end{equation}
where $p_n=|\vp_n|$ is the modulus of $\vp_n$, $\mP_n$ is a rotation matrix
(expressed in the reference frame) 
and $\uz$ is the unit vector of the reference 
frame along the third coordinate. We choose the local frame for the
segment of the path after the $n^{\rm th}$ scattering event and
before the next one. The third coordinate of the local frame is 
therefore always pointing in the direction of motion.

We remark immediately that the decomposition \eqref{eq:decompP} of $\vp_n$
is not unique: The Euler angle $\psi$ of $\mP_n$ can be taken 
arbitrarily, it is the spin gauge freedom. Here stands a 
crucial point in our model: The introduction of the local frame is a natural
way of introducing the spin, and hence the polarization,
into a multiple scattering formulation.
If the wave has a spin
$S$, there are $2S+1$ possible values for the spin component
along the direction of propagation, that we label with ``$s$'' 
($-S\leq s\leq S$). 
At a given point~$\vr$, the field is a superposition of partial fields,
which we consider to be plane waves, with 
different momenta. For simplicity, we may consider 
all directions of propagation for a fixed absolute value of the momentum.
The amplitude of probability for observing the
spin state $s$ in the frame $\mP$ at point $\vr$ and time $t$ is written
$\vPsi(\vr,\,t,\,\mP,\,s)$. In this article, we regularly use 
the bracket notation for functions: 
The functions $f$ of a variable $x$ are written as 
$\Bracket xf$, instead of~$f(x)$.
There is no consequence of this notation concerning the physics itself,
but we find it more convenient for our purpose, in particular for the
introduction of the rotational harmonics in section~\ref{sec:trajstat}.
We then write the amplitude as a bracket product
\begin{equation}
  \vPsi(\vr,\,t,\,\mP,\,s)=\Bracket{\vr,\,\mP,\,s}{\vPsi(t)}.
  \label{def:bracket}
\end{equation}

Polarization is related to the frame of observation and if this
frame is changed, polarization is modified according to certain
rules. Light polarization, for instance, is turned by an angle $-\psi$
if the frame of observation is turned by an angle $\psi$ around the
direction of propagation. It is therefore natural to change the basis
of representation and use the eigenbasis of rotation along the
direction of propagation. Light has two circular polarization
states, corresponding to the helicities of the photon.
A rotation of angle~$\psi$ creates
a phase shift for each circular eigenstate.
If $s$ is an eigenvalue of the spin operator along the direction
of propagation, rotations along this axis commute with the
spin operator. Unitarity implies that
\begin{equation}
  \Bra{\mP\mZ(\psi),\,s}=\e^{\ii s\psi}\Bra{\mP,\,s}
  \label{eq:transfR}
\end{equation}
on the on-shell states.
After these remarks, we shall name the third Euler angle, usually
noted $\psi$, the \emph{spin angle}.

We have presented a new description of the field at a given point
that depends on the local frame and 
thereby taking into account the spin.
We still have to check the completeness of the ket representation
$\Ket{\mP,\,s}$. As we have remarked, the direction of propagation
is $\mP\uz$, so that $\Ket{\mP,\,s}$ and $\Ket{\mP',\,s'}$ are
orthogonal if $\mP\uz\neq\mP'\uz$. From relation~\eqref{eq:decompR}
we conclude that non orthogonal $\Ket{\mP,\,s}$ and $\Ket{\mP',\,s'}$ 
have the same Euler angles except
the spin angle, which is arbitrary.
Let~$\mR$ and~$\mR'$ be two reference frames,
we introduce the function $\deltar(\mR^{-1}\mR')$ to denote
\begin{equation}
  \deltar(\mR^{-1}\mR')
    \equiv4\pi\delta(\cos\theta'-\cos\theta)\delta(\phi'-\phi)
  \label{def:deltar}
\end{equation}
where $\theta$ and $\phi$ are the first two Euler angles of $\mR$
and $\theta'$ and $\phi'$ the first two Euler angles of $\mR'$.
$\deltar$ is a kind of Dirac function for rotation matrices, which
select frames having the same third unit vector.
Using this notation, the condition~$\mP\uz=\mP'\uz$ is imposed
by defining $\Bracket{\mP',\,s'}{\mP,\,s}$ proportional to
$\deltar(\mP^{-1}\mP')$. 
We remark that 
$\deltar(\mP'^{-1}\mP)=\deltar(\mP^{-1}\mP')$.
Finally, the orthogonality of the spin eigenstates
yields $\Bracket{s}{s'}=\delta_{ss'}$.
From the expression~\eqref{eq:transfR} we get the product
\begin{equation}
  \Bracket{\mP',\,s'}{\mP,\,s}= \e^{\ii s'\psi'-\ii s\psi} \;
   \deltar(\mP^{-1}\mP')\;\delta_{ss'}
  \label{eq:scalarRs}
\end{equation}
where $\psi$ and $\psi'$ are the spin angles of $\mP$ and $\mP'$ 
respectively.

The superposition of the partial fields for all 
directions of propagations is expressed by
\begin{equation}
  \begin{split}
  \int_{\SO3}\ud\mP\;\vPsi(\vr,\,t,\,\mP,\,s) \e^{-\ii s\psi}
    &=\frac1{8\pi^2}
      \int_0^\pi \sin\theta\ud\theta 
        \int_0^{2\pi}\ud\phi \int_0^{2\pi}\ud\psi\; 
          \vPsi(\vr,\,t,\,\mZ(\phi)\mY(\theta)\mZ(\psi),\,s)
            \e^{-\ii s \psi}\\
    &=\frac1{4\pi}
      \int_0^\pi \sin\theta\ud\theta \int_0^{2\pi}\ud\phi\; 
        \vPsi(\vr,\,t,\,\mZ(\phi)\mY(\theta),\,s)
          \equiv \vPsi_s(\vr,\,t).
  \label{eq:superp}
  \end{split}
\end{equation}
The presence of the term $\e^{\ii s\psi}$ is essential.
Its role is to add the amplitudes
in such a way that the amplitudes in two distinct frames with the same 
direction of propagation are in phase and do not cancel out.
Forgetting this term would make all $\vPsi_s$ vanish except
$\vPsi_0$, because the exponential term from~\eqref{eq:transfR} would
be integrated to give~$0$. Multiple scattering theories for scalar waves
actually only consider the term~$\vPsi_0$.
The completeness of the Hilbert basis $\Ket{\mP,\,s}$
is expressed by the closure formula ($\opI$ is the identity operator)
\begin{equation}
  \sum_{s=-S}^S \int_{\SO3}\ud\mP\;\Ket{\mP,\,s}\Bra{\mP,\,s} = 
   \opI.
  \label{eq:fermeture}
\end{equation}
The integral over~$\SO3$ has been defined in Eq.~\eqref{eq:superp}.
The description we have introduced in this section will be used in
the formulation of the multiple scattering for polarized waves.
In the next section we introduce the Green's operator for the
polarized states $\Ket{\mP,\,s}$ and its space dependence
in the absence of scatterers.

\section{Transport in a homogeneous medium}
\label{sec:transport}
Before studying multiple scattering, it is necessary to investigate
the transport in a homogeneous medium. By the
word transport, we mean the response at a position $\vr'$ and time $t'$
in the frame $\mP'$ (see previous section for the introduction of
the frames) to a source at position $\vr$ and time $t$ in the frame $\mP$.
It is advantageous to use the Green's functions
because the two situations of massive and massless particles can both
be handled with the same formalism. In the case of a massless particle,
like the photon, transport follows from Maxwell's equations and is often
formulated by the Helmholtz equation.  
Massive particles transport is
of a different nature and the dynamics of their wavefunction
responds to the Schr\"odinger equation. 

We denote  by $\mathcal G$ the Green' function of the 
operator~$\frac{\partial^2}{\partial\vr^2}+\frac{\omega^2}{c^2}$ or of 
the operator
$-\frac{\hbar^2}{2m}\frac{\partial^2}{\partial\vr^2}-E$ to
describe the transport of
the electromagnetic field or of the wavefunction respectively.
The equivalence between these two operators is obtained through
the relation
\begin{equation}
\frac{\omega}{c}=\frac1\hbar\sqrt{2mE},
\label{eq:equivalence}
\end{equation}
so that we shall use in our formul\ae{} only the ``massless''
notation~$\omega/c$.
The results we present can be extended to massive particles of
arbitrary spin by using equation~\eqref{eq:equivalence}
and changing the sign of~$\mathcal{G}$.
If the medium is invariant under translation and in time
$\mathcal{G}(\vr,\,t;\,\vr',\,t')$ depends on $\vr'-\vr$ and $t'-t$. 
In this section, 
we construct a Green's operator~$\opG_0$ which depends on the direction of
motion at~$\vr$ and $\vr'$, and we call it
the directional free Green's operator. 

\begin{figure}
  \includegraphics[width=\textwidth]{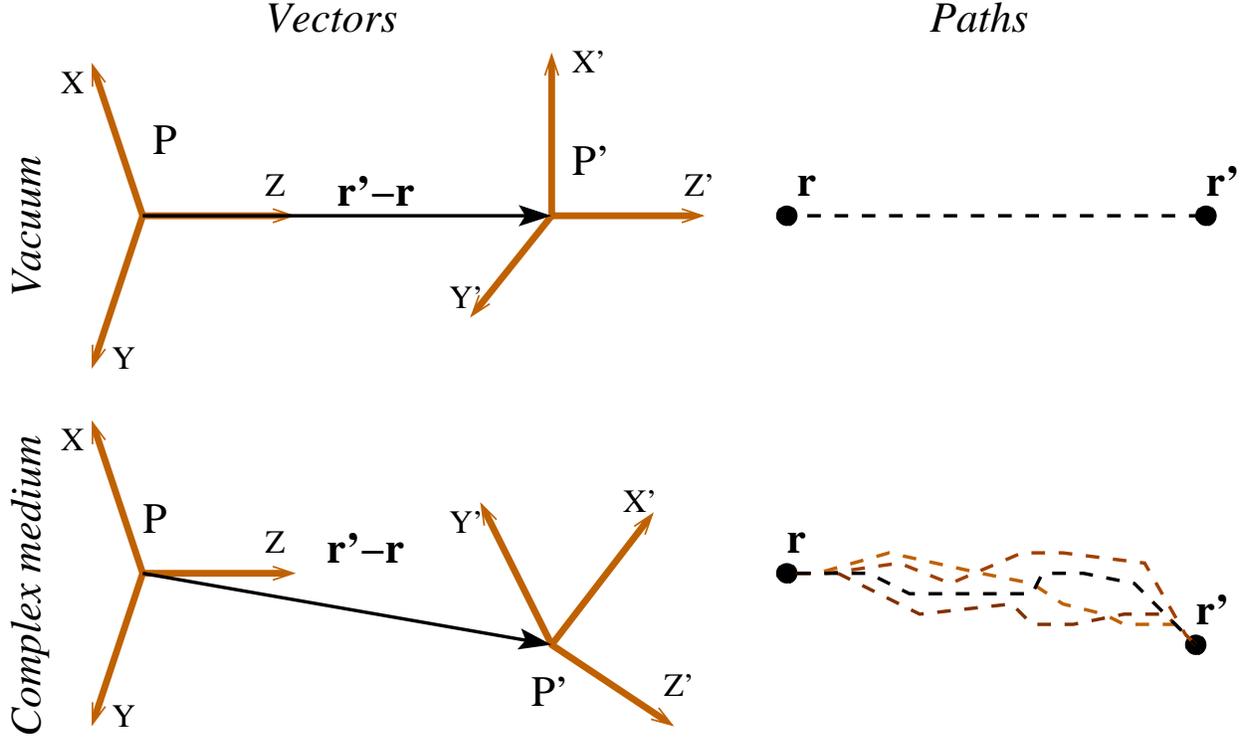}
  \caption{\label{fig:green} Schematic view of the 
  directional Green's operator. Without scatterers (upper figures)
  the only situation for which the Green's operator does not vanish
  is when the three vectors~$\vr'-\vr$, $\mP\uz$ and $\mP'\uz$ are colinear
  and have the same direction (l.h.s. of the figure). 
  In this situation, only one path is possible: The straight
  line from~$\vr$ to $\vr'$ sketched as a dashed line on the r.h.s.
  of the figure.\\
  In a complex medium (lower figures) the directional Green's operator
  $\Bra{\vr',\,\mP'}\opG_0\Ket{\vr,\,\mP}$ describes the transport from
  $\vr$ to $\vr'$ with the constraint that the initial and final
  directions of motion are~$\mP\uz$ and $\mP'\uz$ respectively.
  There are several different paths, some examples of which are shown as
  dashed lines on the r.h.s in the figure. Note that the 
  vectors~$\vr'-\vr$, $\mP\uz$ and $\mP'\uz$
  need not to be colinear anymore.}
\end{figure}

We introduce the free Green's operator as the operator transforming the
wave function along the propagation of the wave if no scattering events occur.
The transition amplitudes characterizing the response may 
depend on the spin.
If we note $\opG_0$ the free Green's operator, 
transport is described by the matrix elements
$\Bra{\vr',\,\mP',\,s'}\opG_0(t',\,t)\Ket{\vr,\,\mP,\,s}$.
We have introduced the position ket $\Ket{\vr}$ and denoted 
$\Ket{\vr,\,\mP,\,s}\equiv\Ket{\vr}\otimes\Ket{\mP,\,s}$.
Naturally, it is not physical in quantum mechanics to consider
the position and the direction of motion of a particle simultaneously.
The directional Green's operator that we need for
our theory can be seen as an intermediate element of computation: The physical
Green's function is the superposition of the directional Green's operators 
for all initial and all final directions of motion. In Feynman's picture
of path integrals, it corresponds simply to decompose the path integral
formulation of~$\mathcal G$ into path integrals over trajectories
with a constraint on the direction of motion at the initial and final point.
We have illustrated this feature in figure~\ref{fig:green}.

In a medium with translational invariance, the momentum is conserved so that 
$\opG_0$ has a factor $\Bracket{\mP}{\mP'}$.
Between two scattering events, the wave travels in space
from $\vr$ to $\vr'$ along the direction $\hat\vp=\mP\uz$. 
The direction of the momentum, $\mP\uz$ is the same as
the direction $\vr'-\vr$ because of the relation
$\vr'-\vr=c(t'-t)\mP\uz$, so that we have to impose
this constraint to the free Green's operator.
To express the directional constraint on~$\vr'-\vr$, 
we write a similar equation
as~\eqref{eq:decompP}~:
\begin{equation}
  \vr'-\vr=r\mD\uz
  \label{eq:decomprr}
\end{equation}
so that the dependence of the Green's operator~$\opG_0$ on~$\mD$ 
is simply reduced to $\deltar(\mD\mP^{-1})$.
Finally, we have the expression for the dynamics Green's operator
\begin{equation}
  \Bra{\vr',\,\mP',\,s'}\opG_{\text{dyn}}(t',\,t)\Ket{\vr,\,\mP,\,s}
  =\mathcal{G}(\vr,\,t;\,\vr',\,t')\, \deltar(\mD\mP^{-1}) \,
    \Bracket{\mP',\,s'}{\mP,\,s}.
  \label{eq:Gdyn}
\end{equation}
The factor $\deltar(\mD\mP^{-1})$ in expression~\eqref{eq:Gdyn}
is a very strong restriction imposed by rotational invariance.
It states that a rotation of the medium must be accompanied by the
same rotation of the frames $\mP$ and $\mP'$ 
(through the term $\Bracket{\mP',\,s'}{\mP,\,s}$) to leave the Green's
operator unchanged. 

In the presence of scatterers, 
it has been observed in backscattering configurations
the transport of polarization depends on 
the relative directions of the incident polarization
beam and the vector~$\vr'-\vr$~\cite{hielscher1997}.
Moreover, in some cases, like in the presence of
linear birefringence or dichroism, 
the direction of propagation has a strong influence on the transport.
Such complex geometrical dependences will be described by the
dependence on~$\mP$, $\mP'$ and $\mD$ of the generalized Green's
operator. 

We now consider the evolution of the spin amplitudes
during the transport.
In the simple case where the time evolution of the 
spin does not depend on position or time, it is 
described by
\begin{equation}
  \opG_{\text{spin}}(t',\,t)=
      \exp\left[-\frac\ii\hbar \opA_{\text{spin}}\,(t'-t)\right],
  \label{eq:spinevol}
\end{equation}
where~$\opA_{\text{spin}}$ is the evolution operator of
the spin.
If there is no anisotropy associated with the spin and if the
spin is conserved during transport, then the terms 
$\Bra{s'}\opG_{\text{spin}}\Ket{s}$ form a $(2S+1)\times(2S+1)$ 
diagonal unitary matrix.
Effects depending on the spin, like circular birefringence or dichroism,
modify the diagonal terms of $\Bra{s'}\opG_{\text{spin}}\Ket{s}$ such
that it is not necessarily unitary anymore. Consider for instance
a medium with absorption length $\kappa_s^{-1}$
and index $n_s$ for the spin $s$. In this case, we have
\begin{equation}
  \Bra{\vr',\,\mP',\,s'}\opG_{\text{spin}}(t',\,t)
    \Ket{\vr,\,\mP,\,s}=
   \e^{-\ii(n_s\omega -\ii c\kappa_s)(t'-t)}\;\delta^{(3)}(\vr'-\vr)\;
   \Bracket{\mP',\,s'}{\mP,\,s} .
  \label{eq:Gspin}
\end{equation}
In this expression, the index $n_s$ denotes the ratio of wave celerity
for a spin eigenstate $s$ compared to the celerity contained in
$\mathcal{G}$. Dichroism appears when the values of $\kappa_s$
depend on~$s$, birefringence when the values of~$n_s$ depend on~$s$.
It is also possible to introduce spin-flips.

To summarize our construction of the Green's operator of
the homogeneous medium we have the
following formula for~$\opG_0$ 
\begin{equation}
  \Bra{\vr',\,\mP',\,s'}\opG_0(t',\,t)\Ket{\vr,\,\mP,\,s}
  = \mathcal{G}(\vr'-\vr,\,t'-t) 
   \BraOpKet{s'}{\opG_{\text{spin}}(t',\,t)}{s}
    \deltar(\mD\mP^{-1})
    \deltar(\mP^{-1}\mP')
    \;\e^{\ii(s'\psi'-s\psi)}.
  \label{eq:Green}
\end{equation}

The dependence of the Green's function as a function of the distance
is, according to the expression~\eqref{eq:Green}, the product of
the scalar Green's function $\mathcal{G}$ and the Green's function
of the spin $\opG_{\text{spin}}$. 
If there is no absorption, the Green's function 
follows the well known $r^{-1}$ decrease
of the Green's function in three dimensions. In this situation, the
enhanced Green's function~\eqref{eq:Green} verifies the conservation
of energy by construction.

\section{Description of a scattering event}
\label{sec:scattering}
In the previous section we have constructed the Green's function
for a homogeneous medium. In this section, we include the description of the
scatterers into the same formalism. The problem of scattering
with spin particles was originally discussed by Jacob and 
Wick~\cite{jacob1959} when two particles with spins are colliding.
We consider more general scatterers in our formalism and
describe an arbitrary interaction in the far-field.
In this situation, we only need to know the so-called on-shell
``$\mT$-matrix'', or transition matrix,
in a theory of multiple scattering.

We use the following definition of the on-shell $\mT$-matrix, in the
case of light
scattering: Consider an incoming field~$\vE_{\text{in}}$ and
a scatterer at position~$\vr$. 
At a far-field position, the scattered field~$\vE_{\text{sca}}$ depends
linearly on $\vE_{\text{in}}$
\begin{equation}
  \vE_{\text{sca}}(\vr',\,t')=
     \mathcal{G}(\vr',\,t';\,\vr,\,t)\mT\vE_{\text{in}}(\vr,\,t).
  \label{def:T}
\end{equation}
The $\mT$-matrix can be computed or measured in several ways
that we will not discuss. 
In general, the $\mT$-matrix depends on the incoming and outgoing directions
and on the polarization. These dependences will be 
rigorously taken into account in the present formalism.

\begin{figure}
\begin{center}
\includegraphics[width=0.48\linewidth]{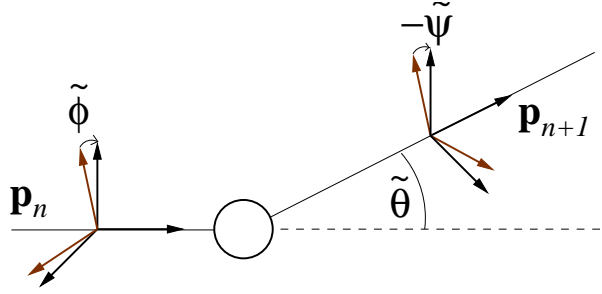}
\caption{\label{fig:scattering}
A single scattering event.
The freedom of rotation of the incoming and outgoing frames is
compensated by a local rotation of the spin angle \eqref{eq:transfR}.
The azimuth angle~$\tilde\phi$ of $\mP_n^{-1}\mP_{n+1}$ determines the
scattering plane and 
the zenith angle~$\tilde\theta$ of $\mP_n^{-1}\mP_{n+1}$ is
the scattering angle.
}
\end{center}
\end{figure}

The position in space of the scatterers
are denoted by $\vr_i$. A scattering event corresponds to a
change in the direction of propagation at $\vr_i$ from $\mP_n\uz$
to $\mP_{n+1}\uz$. In the laboratory frame, the rotation matrix
corresponding to a scattering event is $\mR_0=\mP_{n+1}\mP_n^{-1}$.
However, it is the \emph{local} rotation which is physically
relevant in a local description of scattering. The expression
of this local rotation is $\mR=\mP_n^{-1}\mR_0\mP_n=\mP_n^{-1}\mP_{n+1}$.
This is the local rotation experienced from the point of view of
the particle during the scattering event.
Using the decomposition of $\mR$ into Euler angles ($\tilde\phi,\,\tilde\theta,\,\tilde\psi)$
(Eq.~\eqref{eq:decompR}), we rewrite
$\mP_{n+1}=\mP_{n}\mR$ as
\begin{equation}
  \mP_{n+1}\;\mZ(\tilde\psi)^{-1}=\mP_{n}\;\mZ(\tilde\phi)\mY(\tilde\theta).
  \label{eq:localrot}
\end{equation}
$\tilde\theta$ is the scattering angle and~$\tilde\phi$ and~$-\tilde\psi$
will act as spin angles in~\eqref{eq:transfR}.
As a consequence of this relation, the modifications of the spin angles
of $\mP_{n}$ and $\mP_{n+1}$ transform according to
\begin{gather}
  \mP_{n}\to\mP_{n}\mZ(\tilde\phi)\\
  \mP_{n+1}\to\mP_{n+1}\mZ(-\tilde\psi)
  \label{eq:transfP}
\end{gather}
and cast the rotation into local frames adapted to the
scattering event (see figure \ref{fig:scattering}).

If there is a spherical scatterer at position~$\vr$,
the scattering operator between incoming and outgoing states is
\begin{equation}
\Bra{\vr',\,\mP_{n+1},\,s'}\opT(t',\,t)\Ket{\vr,\,\mP_{n},\,s}=
\delta^{(3)}(\vr'-\vr)
\delta(t'-t) \e^{\ii s\tilde\phi+\ii s'\tilde\psi} 
\langle{\mY(\tilde\theta),\,s'}|\opT\Ket{\mI,\,s}
\label{eq:scattering}
\end{equation}
($\mY$ is a rotation around the $Y$-axis of the reference frame)
We have used the spherical symmetry of the scatterer
to bring the incoming and outgoing
frames back into the incoming one, so that the outgoing frame is simply
$\mY(\tilde\theta)$.
In this formulation, $\langle{\mY(\tilde\theta),\,s'}|\opT\Ket{\mI,\,s}$ is just
the usual $\mT$-matrix written in the spin eigenstates basis, which we write
$\mathcal{T}_{ss'}(\omega,\,\tilde\theta)$.
We impose that there is a scatterer at position~$\vr$ with the
density operator~$\Bra{\vr'}\oprho\Ket{\vr}=\delta^{(3)}(\vr'-\vr)\varrho(\vr)$,
with
\begin{equation}
  \varrho(\vr)=\sum_{i=1}^N\delta^{(3)}(\vr-\vr_i):
  \label{def:rho}
\end{equation}
If there is no scatterer at position~$\vr$ the scattered field vanishes.
The scattered field is obtained using the the closure relation~\eqref{eq:fermeture}
\begin{multline}
  \Bra{\vr',\,\mP',\,s'}\opG_0(t',\,t)\opT\oprho\Ket{\vr,\,\mP,\,s}  \\
 =\int_{\mathcal{V}}\ud^3\vr'' \int_{\SO3}\ud\mP'' \sum_{s''=-S}^{S}
  \Bra{\vr',\,\mP',\,s'}\opG_0(t,\,t)\Ket{\vr'',\,\mP'',\,s''}
  \Bra{\vr'',\,\mP'',\,s''}\opT\varrho\Ket{\vr,\,\mP,s} \\
 =\mathcal{G}(\vr'-\vr,\,t'-t) \deltar(\mD\mP'^{-1}) \varrho(\vr)
  \sum_{s''=-S}^S\Bra{s'}\opG_{\rm spin}(t'-t)\Ket{s''}\;
  \e^{\ii (s'-s'')\psi'}\e^{\ii s\tilde\phi+s''\tilde\psi}
  \mathcal{T}_{ss''}(\omega,\,\tilde\theta).
  \label{eq:scattered}
\end{multline}
To obtain the last equality, we have used the fact that the spin angle of the
rotation matrix~$\mP''$ is arbitrary and does not play any role in the final
expression and we have therefore used the spin angle of~$\mP'$ to simplify
the formula. The angles~($\tilde\phi,\,\tilde\theta,\,\tilde\psi)$ are
the Euler angles of~$\mP^{-1}\mP'$.

The~$\opT$ operator in formula~\eqref{eq:scattered} stands for
a single spherical scatterer, 
correlations or the influence of the other scatterers
are, so far, ignored. 
The imaginary part of~$\opT$ stands for extinction.
The extinction cross-section for spin $s$ is then
\begin{equation}
  \sigma_{\text{ext}}^s=-\frac{c}{\omega}\,\Im\mathcal{T}_{ss}(\omega,\,\theta=0).
  \label{eq:sigmaext}
\end{equation}
It is constructive to compare $\sigma_{\text{ext}}$ with the
scattering cross section for spin $s$
\begin{equation}
  \sigma_{\text{scatt}}^s=\frac12\sum_{s'=-S}^S\int_0^\pi \sin\theta\,\ud\theta
    \left| \mathcal{T}_{ss'}(\omega,\,\theta)\right|^2.
  \label{eq:sigmascatt}
\end{equation}
The total extinction and scattering cross sections are obtained 
by summing over the spin eigenstates: 
$\sigma_{\text{ext}}=\sum_s\sigma_{\text{ext}}^s$ and
$\sigma_{\text{scatt}}=\sum_s\sigma_{\text{scatt}}^s$.
By conservation of energy
the ratio $a=\sigma_{\text{scatt}}/\sigma_{\text{ext}}$ cannot 
exceed~ $1$. The ratio $a$ is called the \emph{albedo}. If
$\sigma_{\text{ext}}=\sigma_{\text{scatt}}$,
all energy captured by the particle is scattered.
This is known as the \emph{optical theorem}. 
Let us remark that the inequality $a\leq1$ valid for the total albedo is 
not necessarily true for the albedoes
$a^s=\sigma_{\text{scatt}}^s/\sigma_{\text{ext}}^s$ at fixed spin.

\section{The Berry phase}
\label{sec:berry}
At this point of the discussion, it is interesting to point out
that our description of the trajectory is able to keep track of 
the Berry phase of the wave. Originally, the Berry phase
was proposed as the phase factor that can appear after a cycle during 
an adiabatic time evolution of a non-degenerated quantum 
state~\cite{berry1984}. Later it was discovered that
the concept applies to light polarization as well~\cite{tomita1986,kwiat1991}
and that the time evolution of the system 
does not need to be cyclic or unitary~\cite{samuel1988}.
More recently, it was shown that the Berry phase also shows up in 
multiply scattered light~\cite{maggs2001d}.
The expression for the Berry phase~$\Phi$ 
involves two factors: The spin~$s$ of
the particle and a solid angle~$\Omega$ (which we call
the geometric phase), or more
generally the curvature of the phase space of the system
enclosed by the evolution of the system during one cycle.
The Berry phase expression is~\cite{berry1984}
\begin{equation}
  \Phi=-s\Omega.
  \label{def:berry}
\end{equation}
In polymer physics, time is replaced by a curvilinear 
space coordinate and the geometric phase~$\Omega$ is 
called~\emph{writhe}~\cite{fuller1971}. The fluctuations
of the writhe induced by thermal fluctuations of the polymer's
shape have been studied numerically using a Monte-Carlo
approach~\cite{rossetto2002a}. For long polymers, the distribution
of the Berry phase was found Gaussian.
There is no exact approach for the statistics of the Berry phase
for random paths.

For simplicity, we consider the equivalent of a
nearly adiabatic time evolution and compare the phases of the
field before and after a path for which 
the initial and final frames are equal. Such a path
corresponds to a closed circuit in the phase space \cite{tomita1986}.
We also assume that the scattering is elastic and that
the spin is conserved in both scattering and propagation. 
Between two scattering events, the 
state of a particle is simply defined by the direction of its
momentum, and the phase space is thus the unit sphere.
The adiabatic evolution would corresponds to a continuous
and slow movement of the direction of the momentum on the sphere.
In case of strong forward scattering
each scattering event corresponds
to a small change in the direction of propagation, which nearly satisfies
the adiabatic condition for the Berry phase to occur. 
The geometric phase is 
the solid angle enclosed by the unit wave vector~$\up=\mP\uz$ along its
trajectory on the unit sphere.

\begin{figure}
  \includegraphics[width=0.5\textwidth]{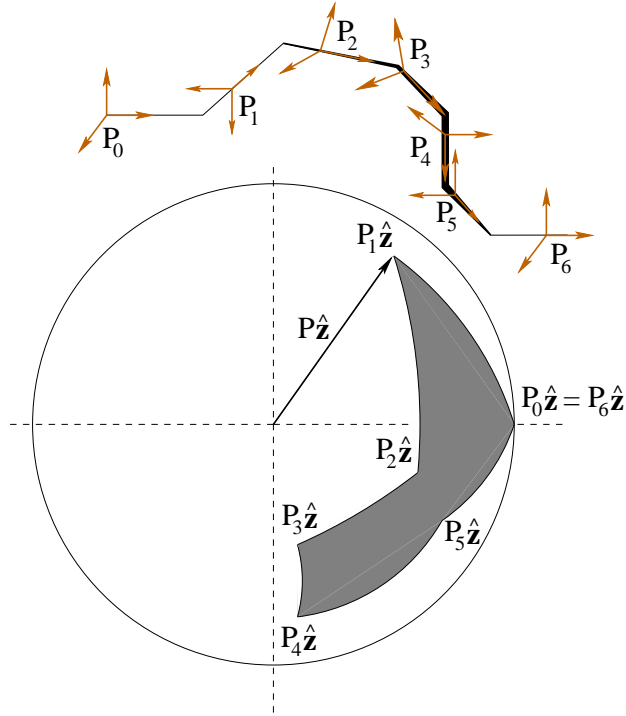}
  \caption{\label{fig:berry} Example in the case~$n=6$ of
  the Berry phase for a trajectory with identical initial and
  final frames. (Top) Trajectory in real space.
  The thickness of the line changes along the trajectory to
  give a three dimensional effect: The closer it is to the observer, 
  the thicker the line is. On each segment is represented a frame
  $\mP_i$ such that $\mP_i\uz$ is the direction of the
  segment. (Bottom) The path is represented on the unit sphere.
  Each segment with constant direction~$\mP_i\uz$ is a point on
  the sphere. Two succesive points are connected by an arc of a circle.
  The area enclosed by the path on the sphere is the geometric phase
  $\Omega$ (in grey). 
  The angles~$\tilde\theta_i$ (see text) have been chosen exaggerately large
  for the readability of both parts of the figure. In strong
  forward scattering, the trajectory on the top part is very close
  to a straight line and the area enclosed on the sphere of the
  bottom part is concentrated around~$\mP_0\uz$. }
\end{figure}

To reveal the Berry phase, 
we consider a trajectory made of $n$ scattering
events, and the~$n+1$ local frames associated to each propagation.
We denote by $(\phi_i,\,\theta_i,\,\psi_i)$ the Euler angles
of each frame after the decomposition of equation \eqref{eq:decompP}.
An example of this construction, with $n=6$, is shown on 
Figure~\ref{fig:berry}.
We consider the case of strong forward scattering so that
the coordinates
$(\theta_{i+1},\,\phi_{i+1})$ and $(\theta_i,\,\phi_i)$ of the
frames $\mP_{i+1}$ and $\mP_i$ are close to one another.
In Figure~\ref{fig:berry}, it corresponds to the case where the
points~$\mP_i\uz$ and $\mP_{i+1}\uz$ on the unit sphere 
are close to each other,
compared to the radius of the sphere.
We call~$(\tilde\phi_i,\,\tilde\theta_i,\,\tilde\psi_i)$
the Euler angles of the rotations~$\mP_i^{-1}\mP_{i+1}$ in the same way 
as we have done in~Eq.~\eqref{eq:localrot}. The strong scattering assumption
corresponds to~$\tilde\theta_i\ll1$. To ensure readability of the 
Figure~\ref{fig:berry}, we have represented a path with large
angles~$\tilde\theta_i$ that do not correspond to strong
forward scattering. 
By analogy with a spinning top, for which the instantaneous
rotation around the third axis is expressed by~$\dot\phi\cos\theta+\dot\psi$
\cite[chap. VI]{landaumeca} we deduce that the rotation around
its direction of propagation experienced by the particle 
is~$\tilde\phi_i+\tilde\psi_i\simeq
(\phi_{i+1}-\phi_i)\cos\theta_i+\psi_{i+1}-\psi_i$.
The Berry phase for a given component of the field is, in our case, 
identified as the phase difference between the values of this
component expressed in the two frames~$\mP_0$ and $\mP_n$. 
The original definition of the Berry phase demands that the evolution
be cyclic. So we have to impose that the frames~$\mP_0$ and~$\mP_n$ are
equal, which corresponds to one cycle. 
Therefore we have $\phi_n=\phi_0\,\mod2\pi$ and $\psi_n=\psi_0\,\mod2\pi$.
The extra phase for a spin state~$s$ between the frames~$0$ and~$n$
is, according to the phase factor~$\e^{\ii(s\tilde\psi+s'\tilde\phi)}$
in equation~\eqref{eq:scattered},
\begin{equation}
  \Phi=s\sum_{i=0}^{n-1}\tilde\phi_i+\tilde\psi_i.
\end{equation}
We have used the conservation of the spin~($s'=s$). We rewrite the sum
using the quantities $(1-\cos\theta_i)$ and 
we get, up to a multiple of $2\pi$:
\begin{equation}
  \Phi=-s\sum_{i=0}^{n-1}(1-\cos\theta_i)(\phi_{i+1}-\phi_i)=-s\Omega
  \quad\mod 2\pi.
  \label{eq:berry}
\end{equation}
We recognize that the sum in~\eqref{eq:berry} is a spherical area.
It is equal to the geometric phase up to a 
multiple of~$4\pi$ because of the indetermination
of the solid angle~\cite{rossetto2002a}. 
We have retrieved the expression of the Berry phase~\eqref{def:berry}.
The equality between the sum in expression~\eqref{eq:berry} and the
geometric
phase holds modulo~$4\pi$, rather than~$2\pi$, for topological
reasons~\cite{rossetto2002b}.
The indetermination modulo~$4\pi$ plays here no role because $s$ is an integer 
or a half-integer, so the phase is undetermined modulo~$2\pi$.

Remarkably, the Berry phase emerges in our local frame model
without any special effort. The geometrical nature of the Berry
phase appears here clearly. As it was already demonstrated,
the existence of the Berry phase is not restricted to cyclic
systems and can be extended to non close path and to non-unitary 
evolution (in our theory, if there is absorption, for instance)
\cite{samuel1988}.
The statistics of the Berry phase for
random walks is a technical and difficult 
problem~\cite{krishna2000,rossetto2005}. 
Our approach will make it possible to obtain exact
results concerning the Berry phase statistics. 

A possible experiment to observe the Berry phase of light consists in
setting a directional lineary polarized source at the edge of a slab of a 
medium filled with scatterers
and observe the outgoing light through a collimator and
a linear polarizer. To ensure strong forward scattering
the scatterers must be large such that
the anisotropy~$g$, defined as the average cosine of the scattering
angle, verifies~$1-g\ll1$.
One can for instance place the polarizer in front of the
source, its plane orthogonal to the direct beam 
and its axis making an angle~$\theta$ with the polarization 
of the source. The intensity as a function
of~$\theta$ will display deviations from the expected $\cos^2\theta$ 
behavior if there were no scatterers. If the slab depth is large,
the Berry phase will spread out on a width larger than~$2\pi$ and
the intensity will be constant as a function of~$\theta$. The experiment
has therefore to be performed with slab depths 
of a few transport mean free paths $\ell^*=\ell/(1-g)$ ($\ell$ is defined
in the next section). 
Different concentrations will lead to different deviations.

\section{Generalized Green-Dyson equation}
\label{sec:trajstat}
The complexity in 
multiple scattering stems from the summation of
the contributions to the field of all the paths from the source 
to the observer, and potentially from their interferences.
In a first approximation one often
neglects these interferences and assumes that the scattering
events are independent from each other. In a disordered system
the averaging over disorder can thus be performed without the
sophisticated diagram techniques needed to preserve interference
effects. The elements of the theory that we have introduced in 
sections~\ref{sec:transport} and \ref{sec:scattering} are used in
this section to write a generalized Green-Dyson equation 
for the multiple scattering of polarized waves under the
assumption of independent scattering events. This equation
applies to directional Green's operators: It relates the
free Green's operator~$\opG_0$ we have 
introduced in section~\ref{sec:transport}
to the effective Green's operator~$\opG$ of the same medium filled
with scatterers. In the second part of this section, we introduce the
Fourier transform of the generalized Green-Dyson equation.

Consider a medium filled with scatterers at fixed positions~$\vr_i$.
The presence of the scatterers is equivalent to the introduction
of the perturbation in the transport equation. 
In the formalism of Green's functions
it corresponds to introducing a transition matrix, or $T$-matrix.
We now consider the perturbed Green's operator~$\opG$
developped as a Born expansion in terms of $\opG_0$:
\begin{equation}
  \opG=\opG_0+\opG_0\opT\oprho\opG_0+\opG_0\opT\oprho\opG_0\opT\oprho\opG_0+\cdots
  \label{eq:Born}
\end{equation}
The first term of the right-hand side of~\eqref{eq:Born} stands
for the unscattered field,
the second term for the single scattering and so on. 

For a known distribution of scatterers, the operator $\opG$ is
the Green's function of the multiply scattering system for polarized waves.
If the functions $\mathcal{G}$, $\mathcal{T}_{ss}(\omega,\,\theta)$ (defined in
section~\ref{sec:transport} and~\ref{sec:scattering} respectively)
and the operator
$\opA_{\text{spin}}$ are known, equation~\eqref{eq:Born} for
$\opG$ solves the problem. But usually we are interested in disordered
systems, for which the exact position of scatterers is unknown.
It will thus be impossible to make a prediction depending on
the particular realization of disorder, but it is possible to
average over all such realizations (ensemble average) and to
write a generalized Green-Dyson equation for the average operators.
We formulate the results using the simplest possible model of disorder.
To perform the average over disorder, we neglect the volume of the
scatterers and consider that each position~$\vr_i$ is uniformly 
distributed in the volume~$\mathcal{V}$. 
The contribution of disorder in~\eqref{eq:Born} comes exclusively 
from the averaging of the operator~$\opT\oprho$.
We introduce the self-energy~$\opSigma(\omega)$
\begin{equation}
  \opSigma(\omega)=\overline{\opT(\omega)\oprho(\vr)}.
  \label{def:Sigma}
\end{equation}
The line denotes the average over all positions for all the
scatterers, which we will explicitely define now for the
calculation of the average density. The average of an 
operator~$\overline\opO$ is simply defined 
for arbitrary~$\Bra{1}$ and~$\Ket{2}$
by~$\Bra{1}\overline\opO\Ket{2}=\overline{\Bra{1}\opO\Ket{2}}$.
If the scatterers all have the same operator~$\opT(\omega)$
the self-energy becomes
\begin{equation} 
  \opSigma(\omega)
    =\overline{\opT(\omega)\oprho(\vr)}
    =\opT(\omega)\int_{\mathcal{V}} \frac{\ud \vr_1}{\mathcal{V}}\cdots
      \int_{\mathcal{V}} \frac{\ud \vr_N}{\mathcal{V}}
      \sum_{i=1}^N \delta^{(3)}(\vr_i-\vr)
    =\opT(\omega)\sum_{i=1}^N \frac1{\mathcal{V}}
    =\opT(\omega)\frac{N}{\mathcal{V}}
    \equiv \rho\opT(\omega),
  \label{eq:moyrho}
\end{equation}
where~$\rho$ is the mean density.
Our approach of disorder is nonetheless naive,
we have neglected the possibility that a trajectory
visits several times the same scatterer and thereby omitted
some possible correlations in the averaging procedure. 
The topic of correlations created by several visits to the same
scatterer is addressed abundantly in the literature. 
The most remarkable result is that for a sufficiently small density of
scatterers, the expression of the Green-Dyson equation remains the
same, but with a modified self-energy
expressed as an expansion in powers of~$\rho$~\cite{vanrossum1999}. 
The first order of the expansion is exactly the expression we have found
in formula~\eqref{def:Sigma}, which is an encouraging fact in
favor of an equivalent expansion in our framework. Such an equivalent 
expansion would extend the domain of validity of our results to higher 
densities. We will address this question in our future work.

Let us call $\opGG$ the average Green's operator resulting from
the averaging over disorder of the operator~$\opG$ defined in~\eqref{eq:Born}.
In the case where all scattering events are independent,
the average of the products involving several times~$\opSigma$
are the products of the averages,
this is sometimes called the Twersky approximation~\cite{mischchenko}.
The Born expansion~\eqref{eq:Born} becomes, in this approximation
the series
\begin{equation}
  \opGG=\opG_0+\opG_0\opSigma\opG_0+\opG_0\opSigma\opG_0\opSigma\opG_0+\cdots
  \label{eq:selfconsGG}
\end{equation}

We have so far expressed the functions only in the direct space of
position and frames. Physical problems however are easier to
formulate in the reciprocal space, because convolutions 
become products. We denote by $\Ket{\vq}$ the reciprocal space
representation of $\Ket{\vr}$ and we have the well known contraction
\begin{equation}
  \Bracket\vr\vq=\e^{\ii\vq\cdot\vr}
  \label{eq:bracketFourier}
\end{equation}
which gives, together with the closure relation 
$\int\ud^3\vr\Ket\vr\Bra\vr=\opI$, the fundamental relations for the
Fourier transform
\footnote{There are several presentations of the Fourier transform,
with different normalizations. Here, the normalisation
consistent with our formul\ae{} is 
$\int\ud^3\vq\Ket\vq\Bra\vq=(2\pi)^3\opI$.}.
Similarly, we use the representation in reciprocal time
that has the same properties as the reciprocal
space. 

As the system is invariant along time translations, the
operators $\opT$, $\opG_0$, $\opGG$, noted as $\opO$,
depend on one frequency~$\omega$.
Similarly, in the average medium obtained after integration over disorder,
the translational invariance is restored the operators
only depend on one vector~$\vq$.
We introduce the notation~$\opO(\omega,\,\vq)$ for operators acting
on the Hilbert space made of the elements~$\Ket{\mP,\,s}$, so that
$\Bra{\vq}\opO(\omega)\Ket{\vq}=\opO(\omega,\,\vq)\,\delta^{(3)}(\vq'-\vq)$.
The representation in reciprocal space for rotation and spin
has different properties and is presented in the next section
using an expression equivalent to~\eqref{eq:bracketFourier}.

The Born expansion written using the operators 
$\opSigma(\omega)$, $\opG_0(\vq,\,\omega)$ and
$\opGG(\vq,\,\omega)$ reduces to
\begin{equation}
   \opGG(\vq,\,\omega) = \opG_0(\vq,\,\omega) 
    + \opG_0(\vq,\,\omega)\opSigma(\omega)\opG_0(\vq,\,\omega)
    + \opG_0(\vq,\,\omega)\opSigma(\omega)\opG_0(\vq,\,\omega)
                          \opSigma(\omega)\opG_0(\vq,\,\omega)
    +\cdots
  \label{eq:Dysonkw}
\end{equation}
The operator $\opSigma(\omega)\opG_0(\omega,\,\vq)$ has a norm 
$\simeq\rho\sigma_{\text{ext}}q^{-1}$ and is therefore
smaller than~$1$ for large $q$ or sufficiently small density~$\rho$.
Consequently, the series $\sum_n(\opSigma(\omega)\opG_0(\omega,\,\vq))^n$
converges and we can rewrite the expansion~\eqref{eq:Dysonkw}
in a self-consistent way
\begin{equation}
  \opGG(\vq,\,\omega)=\opG_0(\vq,\,\omega)+
                      \opG_0(\vq,\,\omega)\opSigma(\omega)\opGG(\vq,\,\omega).
  \label{eq:Dysonselfcons}
\end{equation}
Equation~\eqref{eq:Dysonselfcons} 
is a generalized Green-Dyson equation for the directional Green's operators.
The convergence of the series
occurs when $q\ell>1$ where $\ell\equiv1/\rho\sigma_{\text{ext}}$ is the
extinction mean free path. The theory described by the generalized Dyson
equation does not apply to situations where $q\ell\leq1$
because scatterers are not in the far-field of each other anymore.
In the next section, we introduce the rotational harmonics transform on the
variable $\mP$ to solve equation~\eqref{eq:Dysonselfcons}.

In the expression of the free Green's operator~$\opG_0(\omega,\,\vq)$,
Eq.~\eqref{eq:Green}, 
the factor~$\deltar(\mD\mP^{-1})$ modifies
the usual expression of the Green's 
function~$\sim1/(\frac{\omega^2}{c^2}-\vq^2\pm\ii\eta)$,
where the quantity $\eta$ is a infinitesimaly small positive quantity.
Using the identity 
\[ \int\ud^3\vr\,\frac{\e^{\ii \omega r/c}}{4\pi r} \e^{-\ii\vq\cdot\vr}
   \deltar(\mD\mP^{-1})=\frac1{4\pi}\,
     \frac1{\left(\frac{\omega}{c}-\vq\cdot\mP\uz\right)^2} \]
we get the Fourier transform
\begin{equation}
  \Bra{\mP',\,s'}\opG_0^{R,A}(\omega,\,\vq)\Ket{\mP,\,s}=
   \frac1{\left(\frac{\omega}c-\vq\cdot\mP\uz\right)^2\pm\ii\eta}\;
  \BraOpKet{s'}{\opG_{\text{spin}}(\omega)}{s} \Bracket{\mP'}{\mP}.
  \label{eq:TFG0}
\end{equation}
The $R,A$ superscript stands for the retarded and advanced Green's operator.
As we have already mentioned, the scalar theory must be retrieved
if we sum our expressions over $\mP$ and $s$.  
Here we can check that equation~\eqref{eq:TFG0} integrated over $\mP$
and summed up on $s$ yields formally
\begin{equation}
  \sum_{s=-S}^S\int_{\SO3}\ud\mP 
    \frac1{\left(\frac{\omega}c-\vq\cdot\mP\uz\right)^2}
    \e^{\ii s\psi}\vPsi_s(\vq,\,\omega)
  =\frac1{\frac{\omega^2}{c^2}-\vq^2}\,\vPsi_0(\vq,\,\omega).
  \label{eq:Gusuel}
\end{equation}
The regularization by the infinitesimal imaginary term~$\pm\ii\eta$
can be performed on the r.h.s term of Eq.~\eqref{eq:Gusuel} so that
we retrieve the usual Green's function for the field component
with spin~$s=0$. The superposition of all partial directional
fields cancels out because of the phases of all the spins with $s\neq0$.
The final result depends on~$q=|\vq|$~: the directivity
of the directional Green's operator~\eqref{eq:TFG0} is lost. 
We call the expression~$(\frac{\omega}c-\vq\cdot\mP\uz)^{-2}$ the
directional factor of the Green's operator. 

We can perform the same integration as in formula~\eqref{eq:Gusuel}
on each operator~$\opG_0(\omega,\,\vq)$
in formula~\eqref{eq:Dysonselfcons}. 
We retrieve the scalar Green-Dyson equation~%
$\overline G(\omega,\,q)=\mathcal G(\omega,\,q)+
\mathcal G(\omega,\,q)\Sigma(\omega)\overline G(\omega,\,q)$
with
\[ 
   \Sigma(\omega)=
     \frac12\rho\int_0^\pi \sin\theta\ud\theta\; \mathcal{T}_{00}(\omega,\,\theta).
\]

The Green's operator $\opGG$ does not necessarily act separately
on $\Ket{\vr,\,t}$ and $\Ket{\mP,\,s}$. In optics for instance the
Faraday effect in a medium with Verdet constant $V$ creates a phase
shift $sV\vB\cdot\vp$. In this case, provided that
the spin is conserved during propagation,
equation~\eqref{eq:TFG0} is modified
and the Green's operator $\opG_0$ becomes
\begin{equation}
  \Bra{\mP',\,s'}\opG_0^{R,A}(\vq,\,\omega)\Ket{\mP,\,s}=
    \frac1{\left(\frac\omega c+(sV\vB-\vq)\cdot\mP\uz\right)^2\pm\ii\eta}
     \; \Bra{s'}\opG_{\text{spin}}(\omega)\Ket{s}\Bracket{\mP'}{\mP}.
\label{eq:Faraday}
\end{equation}

The separation of the article into two parts to explain
the harmonic transforms is intended only to keep the explanations 
simple. The Fourier transform presented in this
section and the rotational harmonics transform that
we introduce in the next section could have been presented simultaneously.
This is actually mandatory in the case of a
spin-orbit coupling involving $\Ket{\mP,\,s}$ and $\vq$
like the Faraday effect, as illustrated by formula~\eqref{eq:Faraday}.

\section{Rotational harmonics representation}
\label{sec:rotation}
The linear representation of compact Lie groups provides an harmonical
analysis on the group $\SO3$ which is the equivalent of Fourier
series for periodic functions. These harmonical analysis is
different from the Fourier transform
because $\SO3$ is a compact, non-commutative Lie group. 
For our purpose there is no need to know this mathematical theory
in detail.  Some formul\ae{} used in our theory are
presented in the appendix~\ref{sec:harmonics}.

The reciprocal representations are
labelled by a main index $l\geq0$ which can be an integer or a
half integer and two sub-indices $m$ and $n$ taking the values
$-l,\,-l+1,\dots,\,l$. We denote them by $\Ket{\lmn lmn}$ and we have
\begin{equation}
  \Bracket{\mR}{\lmn lmn}=
      \ii^{n-m}\;\sqrt{2l+1}\;\;
 \e^{\ii m\phi}\;d^l_{m,n}(\cos\theta)\;\e^{\ii n\psi},
  \label{def:Wlmn}
\end{equation}
with $\phi$, $\theta$ and $\psi$ the Euler angles of $\mR$.
The closure relation,
\begin{equation}
  \sum_{l=\Lambda}^\infty
    \sum_{m=-l}^l \sum_{n=-l}^l 
      \Ket{\lmn lmn}\Bra{\lmn lmn} 
    = \opI,
  \label{eq:fermetureW}
\end{equation}
is useful to obtain most of the result provided in the present work.
We call the function $\Bracket{\mR}{\lmn lmn}$ 
the \emph{rotational harmonics}.
(the misleading terminology
``generalized spherical harmonics'' is often used).
These functions are normalized Wigner
$D$-functions\cite{wigner1959,gelfand1956}. 
The indices of the sums on $l$, $m$ and $n$ in 
\eqref{eq:fermetureW} have steps equal to~$1$. 
$\Lambda$, the lowest order of $l$, depends on the spin $S$ 
of the wave and takes the value $0$ or $1/2$ if the spin $S$ is
an integer or a half-integer respectively.
It follows that the indices $l$, $m$ and $n$ are either 
all integers or all half-integers depending on $S$.
For a given $l$ (the order of the harmonics)
there are $(2l+1)^2$ orthogonal functions, which seems a lot.
We will see that for $l\geq S$ only $(2S+1)(2l+1)$ harmonics
components have to be taken into account.
The functions $d^l_{m,n}$ are defined from $ZYZ$ Euler angles
and are real-valued, their full expression is given in 
Equation~\eqref{def:dlmn}. 
We have added the unconventional factor $\ii^{n-m}$ to obey the
identity  $\Bracket{\mR}{\lmn lmn}^*=\Bracket{\mR^{-1}}{\lmn lnm}$
(that is we have suppressed the factor $(-1)^{n-m}$ in this
latter relation).

The rotational dependence of a state can be expressed as a function
of the frame $\mP$ with the bracket notation $\Bracket\mP\vPsi$
or as a function of rotational harmonics indices $\Bracket{\lmn lmn}\vPsi$.
Our computation uses this latter representation because it is related to
the spin in a fundamental way: Let us use the closure relation of the
rotation representation $\int_{\SO3}\ud\mR\,\Ket\mR\Bra\mR=\opI$
to get the expression 
\begin{equation}
  \Bracket{\lmn lmn,\,s}\vPsi= \int_{\SO3}\ud\mR\,
       \Bracket{\lmn lmn,\,s}{\mR,\,s}\Bracket{\mR,\,s}\vPsi.
  \label{eq:decomplmnp}
\end{equation}
According to the relations~\eqref{eq:transfR} and~\eqref{def:Wlmn}
the integration over the spin angle just leaves a factor $\delta_{ns}$,
stating that the representation for a spin eigenvalue $s$ involves
only the kets $\Ket{\lmn lms}$. In the theory of multiple scattering,
$S$ is fixed and there is only one species of particle involved.
It means that in the Hilbert space of physical states the
equality 
$\Ket{\lmn lmn}\otimes\Ket{s}=\Ket{\lmn lms}\otimes\Ket{s}\delta_{ns}$
is valid and the $n$-index can be left out without any 
restriction on the physical content of the theory.

We use the rotational harmonics to expressed the products of the
operators used in the generalized Dyson equation~\eqref{eq:Dysonselfcons}:
\begin{multline}
  \BraOpKet{\lmn {l'}{m'}{s'}}{\opGG^{R,A}(\vq,\,\omega)}{\lmn lms}=
  \BraOpKet{\lmn {l'}{m'}{s'}}{\opG_0^{R,A}(\vq,\,\omega)}{\lmn lms}\\+
  \sum_{\sigma=-S}^S\sum_{L=\Lambda}^\infty\sum_{M=-L}^L
  \sum_{\sigma'=-S}^S\sum_{L'=\Lambda}^\infty
  \sum_{M'=-L'}^{L'}
  \BraOpKet{\lmn {l'}{m'}{s'}}{\opG_0^{R,A}(\vq,\,\omega)}{\lmn LM\sigma}
  \BraOpKet{\lmn LM\sigma}{\opSigma(\omega)}
      {\lmn {L'}{M'}{\sigma'}}
  \BraOpKet{\lmn {L'}{M'}{\sigma'}}{\opGG^{R,A}(\vq,\,\omega)}{\lmn lms}
  \label{eq:DysonSums}
\end{multline}
Would we have used the representation in $\SO3$, we would have
obtained integrals instead of the discrete sums. The sums over $L$
and $L'$ are infinite. However, like in Fourier series, it
is possible to truncate the sum because the brackets amplitudes
decrease for large $L$ like $\alpha^L$ (for a certain $\alpha$, $0<\alpha<1$)
such that high order coefficients can be neglected. 
The sums on indices $L$, $M$ and $\sigma$ can be seen as sums
over a single index. Equation~\eqref{eq:DysonSums} is therefore a
discrete linear equation and is, up to a one-to-one transformation of 
$(L,M,\sigma)$ into a single index, a matrix equation.
If the coefficients of $\opG_0(\vq,\,\omega)$ and
$\opSigma(\omega)$ are known, one can compute $\opGG(\vq,\,\omega)$
with standard linear algebra and get a general solution for the transport
of polarization in multiple scattering by inverting the matrix of
$\opI-\opG_0\opSigma$. 

The directional Green's operator~$\overline\mG$ describes the
transport in a effective, homogeneous medium. Paths statistics
and Berry phases are included in this description, which represent
an improvement over the usual effective Green's functions used
in multiple scattering theories. 
In the presence of several different kinds of scatterers, 
with densities~$\rho_i$
and scattering matrices~$\opT_i$, we get an effective Green's operator
by replacing~$\opSigma$ by $\sum\rho_i\opTT_i$: The scattering
properties of the different scatterers are averaged and their
respective weights are proportional to their respective densities.
If the scatterers are not spherical and have random independent
orientations, the effective~$\opT$-matrix is the average over
all orientations of the orientation-dependent $\opT$-matrices.
In particular, if the orientation probability is uniform, the
average $\opT$-matrix is equivalent to a spherical scatterer's
$\opT$-matrix and the spherical symmetry can be used as is described
in the next section.

To compute the rotational harmonics expansion of~$\opGG$ up to order~$L$,
one needs to use approximately $(2S+1)L^2$ triplets $(l,\,m,\,n)$ 
for the incoming frame,
and as much for the outgoing frame. Therefore required number of 
rotational harmonics coefficients for the computation of the Green's operator 
scales like~$(2S+1)^2L^4$. 
The complexity of the inversion of the linear system~\eqref{eq:DysonSums}
scales approximately like~$L^5$.

\section{The role of rotational invariance}
\label{sec:asym}
We have considered in the previous sections the general
case of an homogeneous medium invariant under translations
and we have decomposed the vectors of the three-dimensional
space into a radius part and an angular part to construct
the general formalism for a directional Green's operator.
A special and important case shows up when the 
medium and the scatterers are also both invariant under rotation.
In the $\SO3$ representation, we have in this case 
for all frames $\mP$ and $\mP'$ the equality already
discussed in section~\ref{sec:scattering}
\begin{equation}
  \Bra{\mP',\,s'}{\opSigma(\omega)}\Ket{\mP,\,s}=
     \Bra{\mP^{-1}\mP',\,s'}{\opSigma(\omega)}\Ket{\mI,\,s}.
  \label{eq:invarT}
\end{equation}
Using the closure relation~\eqref{eq:fermetureW} we get
\[ \Bra{\mP',\,s'}\opSigma(\omega)\Ket{\mP,\,s}=
   \sum_{l=\Lambda}^\infty \sum_{m,n=-l}^l
    \Bracket{\mP^{-1}\mP',\,s'}{\lmn lmn}\Bra{\lmn lmn}
    \opSigma(\omega)\Ket{I,\,s}, \]
We expand the $\Bracket{\mP^{-1}\mP'}{\lmn lmn}$ thanks to 
formula~\eqref{eq:prodW} and sum over~$n$. After
using formula~\eqref{eq:invW}, we obtain
\[ \Bra{\mP',\,s'}\opSigma(\omega)\Ket{\mP,\,s}=
   \sum_{l=\Lambda}^\infty \sum_{m,p=-l}^l \frac1{\sqrt{2l+1}}
   \Bracket{\mP'}{\lmn lp{s'}}
   \Bra{\lmn lm{s'}} \opSigma(\omega)\Ket{I,\,s}
   \Bracket{\mP}{\lmn lpm}.\]
By identification of the terms of the right-hand sum with the rotational
harmonics expansion in $\mP$ and $\mP'$ of the left-hand term%
~\eqref{eq:SerieHarmonique2} we get 
\begin{equation}
  \Bra{\lmn {l'}{m'}{s'}}{\opSigma(\omega)}\Ket{\lmn lms}=
    \frac{\delta_{ll'}\delta_{mm'}}{\sqrt{2l+1}}
   \Bra{\lmn l{s'}{s}}{\opSigma(\omega)}\Ket{\mI,\,s}
   \equiv \frac{\delta_{ll'}\delta_{mm'}}{\sqrt{2l+1}}
   \Sigma\lmn l{s'}s(\omega).
  \label{eq:invarlmn}
\end{equation}
If the medium and the scatterers are both invariant under
rotations, the free Green's operator~$\opG_0$ and
the Green-Dyson operator~$\opGG$ are both independent from
the reference frame. These operators depend on
$\vq$ and on the incoming and outgoing 
directions of propagation~$\mP\uz$ and~$\mP'\uz$,
a rotation of the reference frame acts simultaneously
on these three vectors. Using the decomposition
\begin{equation}
  \vq=q\mQ\uz,
  \label{eq:decompK}
\end{equation}
where $\mQ$ is a rotation matrix and $q=|\vq|$,
we express the invariance of the Green's operators as
\begin{equation}
  \BraOpKet{\mP',\,s'}{\opG^{R,A}(\vq,\,\omega)}{\mP,\,s}=
     \BraOpKet{\mQ^{-1}\mP',\,s'}{\opG^{R,A}(q,\,\omega)}{\mQ^{-1}\mP,\,s},
  \label{eq:invarR}
\end{equation}
where $\opG$ stands either for $\opG_0$ or for $\opGG$.
In Equation~\eqref{eq:invarR}, the 
operator~$\mG$ depends only on $q$ and $\omega$.
Using formula~\eqref{eq:TFG0}, one can even simplify further
the Green's operators by defining
\begin{align}
 \opG_0^{R,A}(q,\,\omega)&=
   \frac{c^2}{\omega^2}\,\opg_0^{R,A}\left(\frac{cq}\omega\right)\\
 \opGG^{R,A}(q,\,\omega)&=\frac{c^2}{\omega^2}\,\opgg^{R,A}
  \left(\frac{cq}\omega\right).
\end{align}
As a consequence of formul\ae{}~\eqref{eq:Green}
we find a symmetry property for~$\opg_0$
\[ \Bra{\lmn {l'}{m'}{s'}}\opg_0^{R,A}\Ket{\lmn lmn}\propto\delta_{mm'}. \]
The same symmetry has been observed for the $\mT$-matrix in 
Equation~\eqref{eq:invarlmn} and we conclude that it will propagate to 
$\opgg$.
A straightforward calculation shows that Equation~\eqref{eq:Dysonselfcons}
takes the form
\begin{equation}
\opgg^{R,A}\left(\frac{cq}\omega\right)=\opg_0^{R,A}\left(\frac{cq}\omega\right)
         +\frac{c^2}{\omega^2} \,
\opg_0^{R,A}\left(\frac{cq}\omega\right)\opSigma(\omega)\,\opgg^{R,A}
\left(\frac{cq}\omega\right).
\end{equation}
The non-vanising rotational harmonics coefficients are related by
\begin{multline}
 \Bra{\lmn {l'}m{s'}}\opgg^{R,A}\left(\frac{cq}\omega\right)\Ket{\lmn lms}
   = \Bra{\lmn {l'}m{s'}}\opg_0^{R,A}\left(\frac{cq}\omega\right)\Ket{\lmn lms}+\\
   \frac{c^2}{\omega^2} \sum_{L=\Lambda}^\infty \frac1{\sqrt{2L+1}}
      \sum_{\sigma=-S}^S \sum_{\sigma'=-S}^S
   \Bra{\lmn{l'}m{s'}}\opg_0^{R,A}\left(\frac{cq}\omega\right)
     \Ket{\lmn Lm{\sigma'}}
   \Sigma\lmn L{\sigma'}\sigma
   \Bra{\lmn Lm\sigma}\opgg^{R,A}\left(\frac{cq}\omega\right)\Ket{\lmn lms}.
   \label{eq:gR}
\end{multline}
Thanks to invariance under rotations, we have obtained a simplified formula
for the generalized Green-Dyson equation. The matrices~$\Sigma\lmn l{s'}s$
are already known for several spherical scatterers, like Mie scatterers.
Formula~\eqref{eq:gR} contains only three sums, two of which, 
over the indices~$\sigma$ and $\sigma'$, are finite sums and correspond
to a matrix product. The expansion of~$\opgg$
into rotational harmonics coefficients is therefore a single sum
over a generalized index instead of a double sum as 
in~\eqref{eq:DysonSums}. The computation
of~$\opgg$ requires to know the rotational harmonics coefficients of the
$\Sigma$-matrix and of~$\opg_0$. The coefficients of $\opg_0$ can be computed 
starting from expression~\eqref{eq:TFG0}~:
\begin{equation}
   \Bra{\lmn{l'}m{s'}}\opg_0^{R,A}(x)\Ket{\lmn lms}=
   \ii^{s'-s}\sqrt{l+\frac12}\sqrt{l'+\frac12}
   \int_{-1}^1\ud\mu\,\frac{d^l_{ms}(\mu)d^{l'}_{ms'}(\mu)}{(1-\mu x)^2
   \pm\ii\eta}.
 \label{eq:g0}
\end{equation}
For arguments $x>1$ there is a pole at $1/x$ in the integral~\eqref{eq:g0}
which has the same physical origin as the pole of the Green's function
in a scalar representation.

\section{Conclusion}
\label{sec:discussion}
We have presented a model for the transport of a polarized wave in a complex
medium containing anisotropies. In our approach, the
wave is represented in the particle picture of quantum mechanics; 
the transport of the wave is described as a superposition of
trajectories with attached probabilities. The polarization of the wave
follows from the spin of the particles. We express the amplitudes
of the spin eigenstates in local frames. 
A local frame is a rotation matrix which third axis is colinear with the
momentum of the particle. Local frames form a convenient setting
for polarized waves because the phases of the spin eigenstates 
of a particle are related to each other by frame dependent factors.
Based on the local frame representation, 
we have established a theory for the multiple scattering of
particles with spin 
and we have obtained a generalized Green-Dyson
equation which takes into account the three dimensional nature of the
transport. One can realize the three-dimensional nature of the 
formalism by noticing that the Berry phase, by essence a three-dimensional
concept, naturally emerges from our equations.

We have suggested to solve the generalized Green-Dyson equation 
using rotational harmonics, which transforms the convolutions
on $\SO3$ into matrix products. The solutions of the generalized 
Green-Dyson equation are obtained by linear algebra operations.
The local frame description is able to take into account several
circular anisotropies, like birefringence and dischroism,
Faraday effect and anisotropic scattering. 
In the special and important case where the scatterers and
the medium are both invariant under rotations, the generalized Green-Dyson
equation takes a simpler form and should be solvable with reasonably light
numerical power. Linear birefringence and dichroism
can also be described in the theory but require a
more complex mathematical treatment which was not
presented here for the sake of simplicity.
Beyond the theory of multiple scattering, we believe that
the generalized Green-Dyson 
operator is a useful object for the statistical study of random 
three-dimensional paths, thanks to its ability to perform path
integrals of direction dependent functionals.

The computation of the solution of the generalized Green-Dyson equation 
using rotational harmonics is probably
the most efficient, because it reproduces the properties of 
the Fourier transform 
used to solve the usual Green-Dyson equation.
On one hand, as the rotational harmonics form
a discrete basis of the algebra on~$\SO3$, one only needs to compute
for each value of $cq/\omega$ a discrete set of coefficients.
On the other hand, the maximum order~$L$ one has to use to get a
physically relevant result grows with the strongest anisotropy
of the system. Despite the simplicity of the
expression of the solutions, it is not conceivable to draw the
calculations analytically in general. The numerical difficulties
lie in two points: The inversion of the linear system~\eqref{eq:DysonSums} 
or~\eqref{eq:gR} and the expression of the operator~$\opg_0$~\eqref{eq:g0}.

We have neglected the interactions between scatterers, which
can create correlations between their positions. Interactions
can be handled by the introduction of the pair correlation function
(or a structure factor) which would enter as a factor in the second
order term of the Born 
expansion\cite{fraden1990}. 
In the terms of higher order~$n$ of the Born expansion, one should
manipulate $n$-points correlation functions, which introduce a 
supplementary difficulty. Much less is known concerning
the interacting scatterers than concerning the self-energy.
We also considered that the medium is globally invariant under
translation and that the density of scatterers is homogeneous,
which excludes small size systems and leads out the effects of boundaries.
Our theory concerns the bulk of a system, but experiments are usually
carried out from the edges. It is therefore an important issue to 
investigate the role of the boundaries.

As such, our work should rather be considered as a first step towards a
full theory of polarization transport in presence of anisotropies.
It may nonetheless already lead to experimental applications and 
investigations in complex media. 
On a more general point of view, our work suggests that polarized 
waves in complex media contain more information concerning anisotropies
than scalar waves and that polarization is therefore a relevant
supplementary observable for the investigation of anisotropic systems.
Our future work will be dedicated to improve the domain of application
of the presented framework to denser, interacting
or quantum systems. First of all, the expression of the self-energy
expansion in powers of the density should be extended up to higher orders
and the formalism for intensity diagrams should be developed.

\section*{Acknoledgments}

The author is indebted to T. Champel,
S. Skipetrov and B. van Tiggelen for their sharp comments
and productive suggestions concerning this work.
This work was partially supported by the ANR 08-JCJC-0066 SISDIF
grant.
 
\appendix
\section{The rotational harmonics}
\label{sec:harmonics}
The rotational harmonics analysis is similar to the Fourier analysis
for periodic functions. It is the direct application of the theorem
of Peter and Weyl~\cite{peter1927} to the representations of
the group $\SO3$. 
Peter-Weyl theorem states that any complex function $f$ defined on 
$\SO3$ may be developped into rotational harmonics series.
The representations of $\SO3$ were introduced in physics by 
Wigner~\cite{wigner1959} but
remain confidential in physical theories mostly because in spinless theories
the rotational harmonics simplify to spherical harmonics. 
A important application in physics was to establish
Wigner-Eckart's theorem~\cite{eckart1930,wigner1959}.
(One can notice that the equation~\eqref{eq:invarlmn} is a particular
case Wigner-Eckart's theorem).
It was shown in section~\ref{sec:rotation} that
the set of rotational harmonics useful in the problem of scattering
of spin particles depends on the spin $S$. More precisely, it
depends whether $S$ is an integer or a half odd integer. 
The rotational harmonics with a half odd integer main index stand for
fermions while the ones with an integer main index stand for
bosons.

The rotational harmonics are given by formula~\eqref{def:Wlmn}
and 
{\small
\begin{equation}
  d^l_{mn}(\cos\theta)=\sqrt{(l-m)!\,(l+m)!\,(l-n)!\,(l+n)!}
    \sum_{p=\max\{0,\,m-n\}}^{\min\{l+m,\,l-n\}}
    (-1)^p\frac{\left(\cos\frac\theta2\right)^{2l+m-n-2p}
                \left(\sin\frac\theta2\right)^{n-m+2p}}
               {(l+m-p)!\,p!\,(n-m+p)!\,(l-n-p)!}.
  \label{def:dlmn}
\end{equation}
}
Remark that the arguments of factorials and the powers are
integers, in both cases where 
$l$, $m$ and $n$ are integers or half-integers.
The computations of the set of $d^l_{mn}$ functions is simplified by
the symmetries 
\begin{align}
d^l_{m,n}(\cos\theta)&=(-1)^{l+m} d^l_{m,-n}(-\cos\theta)
                      =(-1)^{l+m} d^l_{m,-n}(\cos(\pi-\theta)),\\
                     &=(-1)^{m-n} d^l_{-m,-n}(\cos\theta),\\
                     &=(-1)^{m-n} d^l_{n,m}(\cos\theta).
\end{align}
One can also define the rotational harmonics from the spherical 
harmonics by mean of an operator introduced in references~\cite{newman1966}.
This result is sketched in the next appendix.

In our notations the harmonics expansion is written
\begin{equation}
  f(\mR)=\Bracket\mR f=
  \sum_{l\geq\Lambda}\;\sum_{m,n=-l}^l\;
    \Bracket\mR{\lmn lmn}\Bracket{\lmn lmn}f,
  \label{eq:SerieHarmonique1}
\end{equation}
where the harmonics coefficients are 
\begin{equation}
  \Bracket{\lmn lmn}f=\int_{\SO3}\ud\mR\;\Bracket{\lmn lmn}\mR\Bracket\mR f
                     =\int_{\SO3}\ud\mR\;\Bracket{\mR}{\lmn lmn}^*\,f(\mR).
  \label{eq:TermeHarmonique1}
\end{equation}
Similarly, the operators can be expressed in the basis of the rotational
harmonics. 
\begin{gather}
\mO(\mR',\mR)=\Bra{\mR'}\opO\Ket\mR
  =\sum_{l,l'\geq\Lambda}\;\sum_{m,n=-l}^l\;\;\sum_{m'n'=-l'}^{l'}
 \Bra{\lmn {l'}{m'}{n'}}\opO\Ket{\lmn lmn}\Bracket{\mR'}{\lmn {l'}{m'}{n'}}
    \Bracket{\mR}{\lmn lmn}^*,
\label{eq:SerieHarmonique2}\\
\text{where}\;\;
\Bra{\lmn {l'}{m'}{n'}}\opO\Ket{\lmn lmn}=
  \int_{\SO3}\ud\mR'\int_{\SO3}\ud\mR\;
  \mO(\mR',\mR) \; \Bracket\mR{\lmn {l'}{m'}{n'}}^*\;
     \Bracket{\mR}{\lmn lmn}.
\label{eq:TermeHarmonique2}
\end{gather}

The product of operators $\opC=\opA\opB$ is expressed in 
direct space as a convolution
\begin{equation}
  \Bra{\mR'}\opC\Ket\mR=
    \int_{\SO3}\Bra{\mR'}\opA\Ket\mX\Bra\mX\opB\Ket\mR\,\ud\mX
\end{equation}
and in harmonics space as a series
\begin{equation}
  \Bra{\lmn {l'}{m'}{n'}}\opC\Ket{\lmn lmn}=
    \sum_{L=\Lambda}^\infty\sum_{M,N=-L}^L
      \Bra{\lmn {l'}{m'}{n'}}\opA\Ket{\lmn LMN}
      \Bra{\lmn LMN}\opB\Ket{\lmn lmn}.
\end{equation}
The sum over $M$ and $N$ is the usual matrix product. In our calculations
we also use the relations
\begin{gather}
   \sum_p\Bracket{\mR_1}{\lmn lmp}\Bracket{\mR_2}{\lmn lpn}
        =\sqrt{2l+1}\,\Bracket{\mR_1\mR_2}{\lmn lmn}, \label{eq:prodW} \\
 \Bracket{\mR^{-1}}{\lmn lmn}={\Bracket{\mR}{\lmn lnm}}^*. \label{eq:invW}
\end{gather}

It is also interesting to note that we have the 
Parseval-Plancherel formul\ae{}:
\begin{gather}
\Bracket ff=\int_{\SO3}|f(\mR)|^2\ud\mR=
   \sum_{l=\Lambda}^\infty\sum_{m,n=-l}^l \left|\Bracket{\lmn lmn}f\right|^2,
  \label{eq:Plancherel}\\
  \int\ud\mR\int\ud\mR' \; \left|\Bra{\mR'}\opO\Ket\mR\right|^2=
  \sum_{l,l'=\Lambda}^\infty \sum_{m,n=-l}^l \sum_{m',n'=-l'}^{l'} 
  \left|\Bra{\lmn {l'}{m'}{n'}}\opO\Ket{\lmn lmn}\right|^2.
\label{eq:Plancherel2}
\end{gather}

\begin{figure}
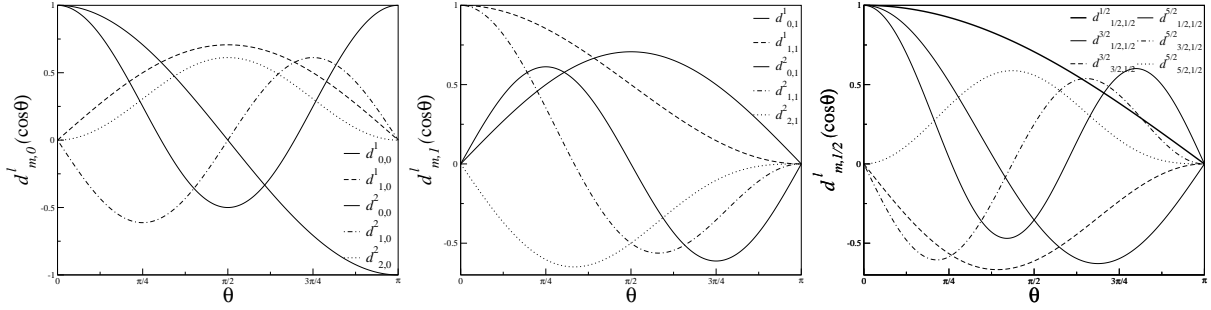

  \begin{center}
    \includegraphics[width=0.32\textwidth]{dlm0.eps}
    \includegraphics[width=0.32\textwidth]{dlm1.eps}
    \includegraphics[width=0.32\textwidth]{dlm12.eps}
  \end{center}
  \caption{\label{fig:dlmn}Examples of functions 
  $d^l_{mn}(\cos\theta)$ for small values of $l$.
  Left: graphs of functions with spin index~$0$ for
  $l=1$ and $l=2$.
  One recognizes the unnormalized spherical harmonics.
  The only function with $l=0$, $d^0_{00}=1$ is not shown.
  Middle: graphs of functions with spin index~$1$
  for $l=1$ and $l=2$.
  Right: graphs of functions with spin index~$1/2$
  for $l=1/2$ and $l=3/2$. }
\end{figure}

\section{Rotational harmonics under other names}
\label{sec:other}
In this appendix we compile the relations between the
rotational harmonics and other special functions used in literature
under several names and with a large variety of normalizations.
This appendix is intended to allow those familiar with one the
numerous forms of the rotational harmonics to comprehend
our presentations from a wider point of view.
Conventionnally, the argument~$\cos\theta$ may be written~$\mu$. 

Wigner introduced the rotational harmonics arranged in matrices called
Wigner's D-matrices of order $l$~\cite{wigner1959}. The elements of
$D(l)$ are given by
\begin{equation}
  D^l_{m'm}(\phi,\,\theta,\,\psi)=\e^{-\ii m'\phi}\,d^l_{m'm}(\cos\theta)
           \e^{-\ii m\psi}.
  \label{def:D}
\end{equation}

\subsection{Functions $\pi$ and $\tau$ used in the Mie expansion}
The well known Mie theory for spherical
dielectric scatterers provides the
general solution of the scattering of a plane wave by a
dielectric sphere of arbitrary size and refraction 
index\cite{mie1908,bornwolf}.
Forward scattering appears for large spheres of radius~$a\gg 2\pi/\lambda$,
this is sometimes called the \emph{Mie effect}.
The scatterers are spherical, so they are invariant under rotations 
and the $\mT$-matrix is thus of the form of Equation~\eqref{eq:invarlmn}.
Written in the spin eigenstates basis (also called
circular polarization basis), the $\mT$-matrix
coefficients take the form
\begin{equation}
  \mathcal{T}_{ss'}(\omega,\,\theta)=\frac{2\ii\pi c}{\omega}
               \ii^s \left(S_1(\theta)+ss' S_2(\theta)\right)
  \qquad s,s'=\pm1
\label{eq:Miecirc}
\end{equation}
with the coefficients $S_1(\theta)$ and $S_2(\theta)$ 
as defined by van de Hulst~\cite{vandehulst1957}.

Mie theory provides the expansion of the coefficients
$S_1$ and $S_2$ 
into series of terms involving special functions referred
as $\pi_n$ and $\tau_n$ in \cite{vandehulst1957} and many
other textbooks. The coefficients of the expansion
are called $a_n$ for $S_1$ and $b_n$ for $S_2$.
In the circular polarization basis, we obtain
the main coefficients as combination of $\pi_n$ and $\tau_n$: 
\begin{equation}
\left\{\begin{array}{ccc}
    S_1(\theta)+S_2(\theta)&=&
    \sum_{n=1}^\infty(2n+1)\frac{a_n+b_n}2
       \frac{\pi_n(\cos\theta)+\tau_n(\cos\theta)}{n(n+1)},\\
    S_1(\theta)-S_2(\theta)&=&
    \sum_{n=1}^\infty(2n+1)\frac{a_n-b_n}2
       \frac{\pi_n(\cos\theta)-\tau_n(\cos\theta)}{n(n+1)}.
\label{eq:Miecirccoeff}
\end{array}\right.
\end{equation}

The functions $\pi_l(\mu)$ and $\tau_l(\mu)$
are defined by 
\begin{gather}
  \pi_0=0,\quad
  \pi_1=1\;\text{and}\quad
  \pi_{l+1}(\mu)=\frac{2l+1}l\mu\pi_l(\mu)-\frac{l+1}l\pi_{l-1}(\mu);\\
  \tau_l(x)=l\mu\pi_l(\mu)-(l+1)\pi_{l-1}(\mu).
\end{gather}
As it was shown by Domke~\cite{domke1975}, we have the relations:
\begin{align}
  \pi_{l}(\mu)&=l(l+1)\left(d^l_{1,1}(\mu)+d^l_{1,-1}(\mu)\right),
  \label{eq:pi}\\
  \tau_l(\mu)&=l(l+1)\left(d^l_{1,1}(\mu)-d^l_{1,-1}(\mu)\right).
  \label{eq:tau}
\end{align}
(In~\cite{newton}, the functions $\pi$ and $\tau$ are
defined with the same equations, without the factor~$l(l+1)$.)

It appears that the Mie expansion is nothing but the 
rotational harmonics expansion 
for spins $s=\pm1$ of a spherical scatterer's $\opT$-matrix.
In the low density approximation used in the article, the self-energy
matrix defined in~\eqref{eq:invarlmn} is then
\[ \Sigma\lmn l{1,}{\pm1}=
-\Sigma\lmn {l}{-1,}{\mp1}=2\ii\pi(2l+1)\frac{\rho c}\omega\;
\frac{a_l\pm b_l}2.\]
One should not be surprised because Mie's result comes
from the resolution of Maxwell's equations, which are vectorial,
thus by nature concerning fields of spin~$S=1$.

\subsection{Vector spherical harmonics}
In electromagnetism, the introduction of spherical harmonics
is motivated by the possibility of multipole expansion to solve
several problems~\cite{jackson,newton}. The multipolar expansion
of a vector field can be expressed in terms of ``vector
spherical harmonics''. There does not seem to be a standard
notation for these vectors. In Ref.~\cite{jackson}, one finds the
definition of~$\mathbf{X}_{lm}(\theta,\,\phi)$ while in Ref~\cite{newton}
we have 
\begin{align}
\mathbf{Y}^{(e)}_{lm}(\vr)&=\frac1{\sqrt{l(l+1)}}r\mathbf\nabla Y_{lm},\\
\mathbf{Y}^{(m)}_{lm}(\vr)&=r\mathbf{X}_{lm}
                           =\hat\vr\times\mathbf{Y}^{(e)}_{lm},\\
\mathbf{Y}^{(o)}_{lm}(\vr)&=Y_{lm}\,\vr.
\end{align}
The spherical harmonics~$Y_{lm}$  being given by
\begin{equation}
Y_{lm}(\theta,\,\phi)=\sqrt{\frac{2l+1}{4\pi}}\,
\sqrt{\frac{(l-m)!}{(l+m)!}}\,
  d^l_{m,0}(\cos\theta)\,\e^{\ii m\phi},
\end{equation}
we have in spherical coordinates 
$(\hat\vr,\hat{\boldsymbol\theta},\hat{\boldsymbol\phi})$
\begin{align}
\mathbf{Y}^{(e)}_{lm}(\vr)=
\frac{(-1)^m\ii^l}{\sqrt{l(l+1)}} 
\sqrt{\frac{2l+1}{4\pi}}
\sqrt{\frac{(l-m)!}{(l+m)!}}
\frac{\e^{\ii m\phi}}{\sin\theta}
\begin{pmatrix}0\\\frac{1-\mu^2}2\left(d^l_{m,1}(\mu)+d^l_{m,-1}(\mu)\right)\\
  \ii md^l_{m,0}(\mu)
\end{pmatrix}.
\end{align}
Only the vector spherical harmonics with~$m=\pm1$ are used to 
solve electromagnetic problems. We find again that it is more natural
to look for the solutions using the basis adapted to spin~$S=1$ in
electromagnetism.

\subsection{Generalized spherical harmonics}
In the work of Ku\v{s}\v{c}er and Ribari\v{c}, ``generalized spherical
harmonics'' have been introduced for the Stokes parameters describing
polarized light \cite{kuscer1959}. Stoked parameters are intensities
and therefore correspond to the amplitude of the wave field squared.
This is why the rotational harmonics appear in this work and followers
with a power of two \cite{domke1975,siewert1982,hovenier1983}. In other
words, the expansion involves the functions~$d^l_{m,0}$ and 
$d^l_{m,\pm2}$. The spherical functions are, in these papers, noted 
$P_{mn}^l(\cos\theta)$ and are simply
\begin{equation}
  P_{mn}^l(\mu)=\ii^{n-m}\,d^l_{mn}(\mu).
\end{equation}
Our rotational harmonics $\Bracket{\mR}{\lmn lmn}$ 
only differ from these spherical functions by a normalization
factor~$\sqrt{2l+1}$ which we introduced to simplify the
expression of the convolution formul\ae.

\subsection{Spin-weighted spherical harmonics ${}_sY_l^m$}
Newman and Penrose introduced functions called the ``spin-weighted
spherical harmonics'' to study gravitational radiations (which have
a spin $S=2$) \cite{newman1966}. The spin weighted functions are
obtained after derivation of the spherical harmonics by the operator
$\thop$ (read ``thop'') defined, for a spin-weight $s$ function $f_s$, by
\begin{equation}
  \thop f_s(\theta,\phi)=-(\sin\theta)^s
     \left[\frac\partial{\partial\theta}+
      \ii\frac1{\sin\theta}\frac\partial{\partial\phi}\right]
      (\sin\theta)^{-s}f_s(\theta,\phi).
\end{equation}
The usual spherical harmonics are of spin-weight~$0$
and the spin-weighted spherical harmonics of spin $s$ are
defined by
${}_sY_{lm}(\theta,\phi)=\frac12\sqrt{\frac{(l-s)!}{(l+s)!}} \thop^s 
Y_{lm}$ for $s>0$ and 
${}_sY_{lm}(\theta,\phi)=\frac12\sqrt{\frac{(l-s)!}{(l+s)!}} (-\bar\thop)^s 
Y_{lm}$ for $s<0$. We have the relation
\begin{equation}
  {}_sY_{lm}(\theta,\,\phi)=(-1)^{m+s}
    \sqrt{\frac{2l+1}{4\pi}}\; d^l_{m,-s}(\cos\theta)\;\e^{\ii m\phi}.
\end{equation}

\bibliography{article}

\end{document}